\documentclass[aps,
prl,
twocolumn,
amsmath,
amssymb,
floatfix]{revtex4-1}

\usepackage{graphicx}
\usepackage{dcolumn}
\usepackage{bm}
\usepackage{dsfont}
\usepackage{physics}
\usepackage{amsmath,amsthm,amsfonts,amssymb,amscd}
\usepackage[mathlines]{lineno}
\usepackage[usenames, dvipsnames]{color}
\usepackage{mathtools}
\usepackage{textcomp}
\usepackage{hyperref}
\hypersetup{
    colorlinks=true,
    linkcolor=black,
    filecolor=black,      
    urlcolor=black,
    citecolor = black
}

\def\u{\ensuremath\uparrow}
\def\d{\ensuremath\downarrow}

\usepackage{bbold}
\usepackage{titlesec}
\titlespacing*{\section} {0pt}{3.5ex plus 2ex minus .2ex}{2.3ex plus .2ex}
\allowdisplaybreaks

\begin{document}
\preprint{}

\title{Observation of a many-body-localized discrete time crystal \\ with a programmable spin-based quantum simulator}

\author{J. Randall$^{1,2}$}\thanks{These authors contributed equally to this work.}
\author{C. E. Bradley$^{1,2}$}\thanks{These authors contributed equally to this work.}
\author{F. V. van der Gronden$^{1,2}$}
\author{A. Galicia$^{1,2}$}
\author{M. H. Abobeih$^{1,2}$}
\author{\\ M. Markham$^3$} 
\author{D. J. Twitchen$^3$}
\author{F. Machado$^{4,5}$} 
\author{N. Y. Yao$^{4,5}$} 
\author{T. H. Taminiau$^{1,2}$}
 \email{t.h.taminiau@tudelft.nl}

\affiliation{$^{1}$QuTech, Delft University of Technology, PO Box 5046, 2600 GA Delft, The Netherlands}
\affiliation{$^{2}$Kavli Institute of Nanoscience Delft, Delft University of Technology, PO Box 5046, 2600 GA Delft, The Netherlands}
\affiliation{$^{3}$Element Six Innovation, Fermi Avenue, Harwell Oxford, Didcot, Oxfordshire OX11 0QR, United Kingdom}
\affiliation{$^{4}$Dept. of Physics, University of California, Berkeley, CA 94720 USA}
\affiliation{$^{5}$Materials Sciences Division, Lawrence Berkeley National Laboratory, Berkeley, CA 94720, USA}

\date{\today}

\begin{abstract}
The discrete time crystal (DTC) is a recently discovered phase of matter that spontaneously breaks time-translation symmetry. Disorder-induced many-body-localization is required to stabilize a DTC to arbitrary times, yet an experimental investigation of this localized regime has proven elusive. Here, we observe the hallmark signatures of a many-body-localized DTC using a novel quantum simulation platform based on individually controllable $^{13}$C nuclear spins in diamond. We demonstrate the characteristic long-lived spatiotemporal order and confirm that it is robust for generic initial states. Our results are consistent with the realization of an out-of-equilibrium Floquet phase of matter and establish a programmable quantum simulator based on solid-state spins for exploring many-body physics.
\end{abstract}

\maketitle

A time crystal spontaneously breaks time-translation symmetry \cite{wilczek2012quantum}. While time crystals cannot exist for time-independent Hamiltonians \cite{watanabe2015absence}, it is predicted that periodically driven `Floquet' quantum many-body systems can break discrete time-translation symmetry \cite{else2016floquet,khemani2016phase,yao2017discrete,khemani2019brief,else2020discrete}. Such a discrete time crystal (DTC) spontaneously locks onto a period that is a multiple of that of the drive and is stable against perturbations. It represents a novel phase of matter that only exists out of equilibrium and exhibits long-range spatial and temporal order. Stabilizing the DTC phase to arbitrary times requires disorder-induced many-body-localization (MBL), which prevents heating from the periodic drive and induces a breakdown of ergodicity \cite{else2016floquet,khemani2016phase,abanin2019colloquium,khemani2019brief,else2020discrete,d2014long,yao2017discrete}. 

Pioneering experiments have revealed signatures of time-crystalline order in a range of systems including trapped ions \cite{zhang2017observation,kyprianidis2021observation}, spin ensembles \cite{choi2017observation,osullivan2020signatures, rovny2018observation,pal2018temporal}, ultracold atoms \cite{smits2018observation, autti2018observation} and superconducting qubits \cite{frey2021simulating}. However, none of these experiments satisfy the theoretical requirements for MBL under periodic driving \cite{khemani2019brief,ippoliti2020many}. The observed responses have instead been attributed to a variety of fascinating critical and prethermal mechanisms that lead to slow, but finite, thermalization \cite{else2017prethermal,khemani2019brief,else2020discrete,machado2020long,peng2021floquet,kyprianidis2021observation,ippoliti2020many}. Experimentally investigating the DTC phase, stabilized by MBL, has remained an outstanding challenge \cite{khemani2019brief,ippoliti2020many}. 

Here, we present an observation of the hallmark signatures of the many-body-localized DTC phase. We develop a quantum simulator based on individually controllable and detectable $^{13}$C nuclear spins in diamond, which can be used to realize a range of many-body Hamiltonians with tunable parameters and dimensionalities. We show that this simulator can be programmed to satisfy all requirements for a DTC, including stabilizing MBL under periodic driving. We implement a periodic Floquet sequence in a one-dimensional (1D) chain of $L = 9$ spins, and observe the characteristic long-lived DTC response with twice the driving period. By combining the ability to prepare arbitrary initial states with site-resolved measurements, we confirm the DTC response for a variety of initial states up to $N = 800$ Floquet cycles. This robustness for generic initial states provides a key signature to distinguish the many-body-localized DTC phase from prethermal mechanisms, which only show a long-lived response for selected states \cite{ippoliti2020many,khemani2019brief, kyprianidis2021observation}. 

\begin{figure*}
    \centering
    \includegraphics[width=1.0\textwidth]{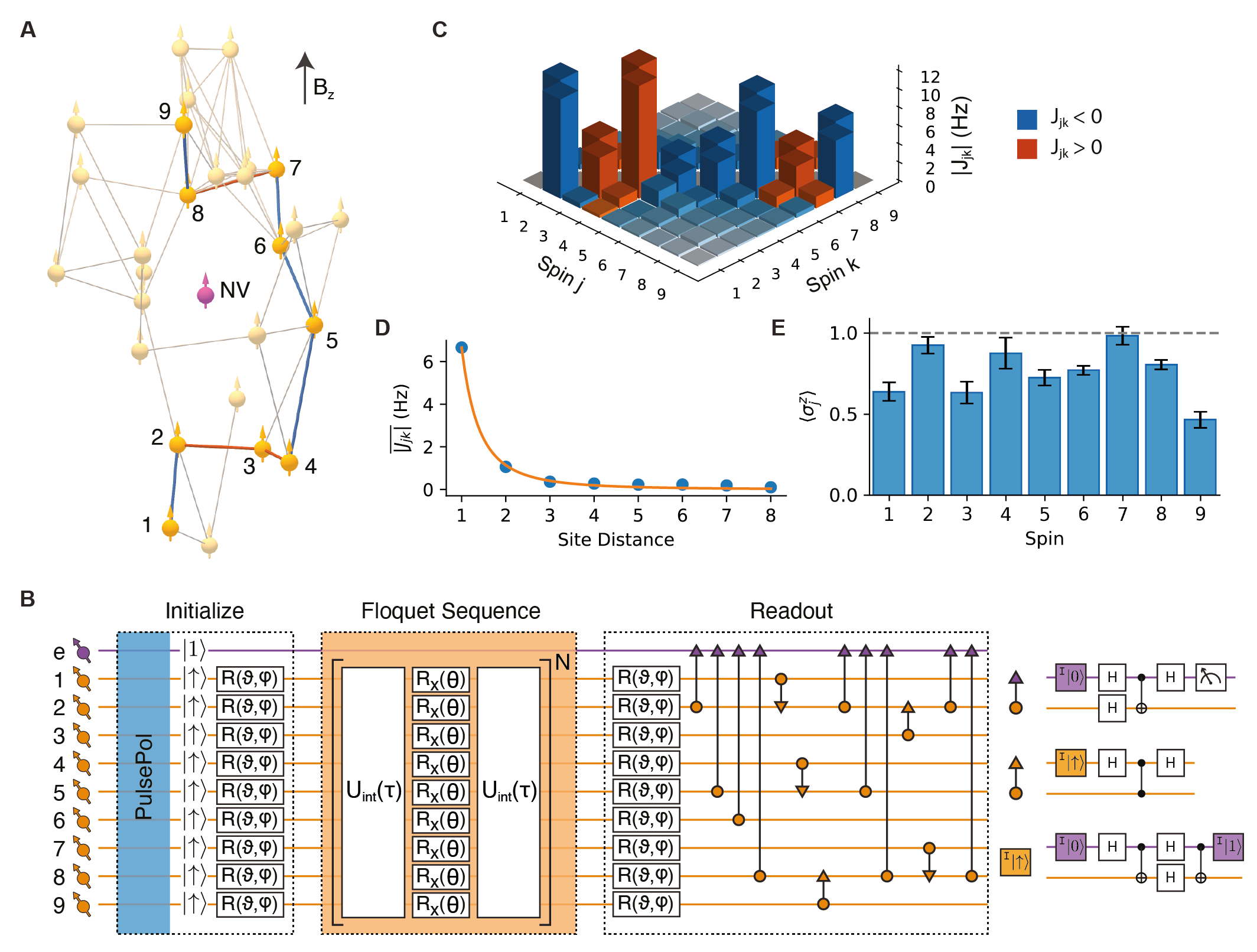}
    \caption{{\bf Programmable spin-based quantum simulator.} ({\bf A}) We program an effective 1D chain of 9 spins in an interacting cluster of 27 $^{13}$C nuclear spins (orange) close to a single NV center. Connections indicate nuclear-nuclear couplings $|J_{jk}|> 1.5\,$Hz, and blue (red) lines represent negative (positive) nearest-neighbor couplings within the chain \cite{abobeih2019atomic}. Magnetic field: $B_z \sim 403\,$G. ({\bf B}) Experimental sequence: The spins are initialized by applying the PulsePol sequence \cite{schwartz2018robust}, followed by rotations of the form $R(\vartheta,\varphi) = \exp[-i\frac{\vartheta}{2} (\sin(\varphi)\sigma^x + \cos(\varphi)\sigma^y)]$. After evolution under the Floquet sequence $U_F = [U_\text{int}(\tau)\cdot U_x(\theta) \cdot U_\text{int}(\tau)]^N$, the spins are sequentially read out through the NV electronic spin using electron-nuclear and nuclear-nuclear two-qubit gates (see text). Colored boxes with `I' denote re-initialization into the given state. ({\bf C}) Coupling matrix for the 9-spin chain. ({\bf D}) Average coupling magnitude as a function of site distance across the chain. Orange line: least-squares fit to a power-law function $J_0/|j-k|^{\alpha}$, giving $J_0 = 6.7(1)\,$Hz and $\alpha = 2.5(1)$. ({\bf E}) Measured expectation values $\langle \sigma_j^z \rangle$ after initializing the state $\ket{\u\u\u\u\u\u\u\u\u}$. The data is corrected for measurement errors \cite{supp}.}
    \label{fig:1Dchain}
\end{figure*}

Our experiments are performed on a system of $^{13}$C nuclear spins in diamond close to a nitrogen-vacancy (NV) center at 4 K (Fig. \ref{fig:1Dchain}A). The nuclear spins are well-isolated qubits with coherence times up to tens of seconds \cite{bradley2019ten}. They are coupled via dipole-dipole interactions and are accessed through the optically addressable NV electronic spin \cite{bradley2019ten, abobeih2019atomic}. With the electronic spin in the $m_s = -1$ state, the electron-nuclear hyperfine interaction induces a frequency shift $h_j$ for each nuclear spin, which --- combined with an applied magnetic field $B_z$ in the $z$-direction --- reduces the dipolar interactions to Ising form \cite{supp}. We additionally apply a radio-frequency (rf) driving field to implement nuclear-spin rotations. The nuclear-spin Hamiltonian is then given by $H = H_\text{int} + H_\text{rf}$, where $H_\text{int}$ and $H_\text{rf}$ describe the interaction and rf driving terms respectively:
\begin{equation}\begin{split}\label{eq:Hint}
    H_\text{int} &= \sum_{j} (B + h_j) \sigma_j^z + \sum_{j<k} J_{jk} \sigma_j^z \sigma_k^z \\
    H_\text{rf} &= \sum_j \Omega(t) \sigma_j^x.
\end{split} \end{equation}
Here $\sigma_j^\beta$, $\beta = x,y,z$ are the Pauli matrices for spin $j$, $B = \gamma_c B_z/2$ is the magnetic field splitting, $\gamma_c$ is the $^{13}$C gyromagnetic ratio, $J_{jk}$ is the $zz$ component of the dipole-dipole interaction between spins $j$ and $k$, $\Omega(t)$ is the applied time-dependent rf field and we set $\hbar = 1$. The system has previously been characterized in detail \cite{abobeih2019atomic, supp}; for 27 $^{13}$C spins the hyperfine shifts $h_j$, the spatial coordinates, and the 351 interaction terms $J_{jk}$ are known.

\begin{figure}
    \centering
    \includegraphics[width=0.5\textwidth]{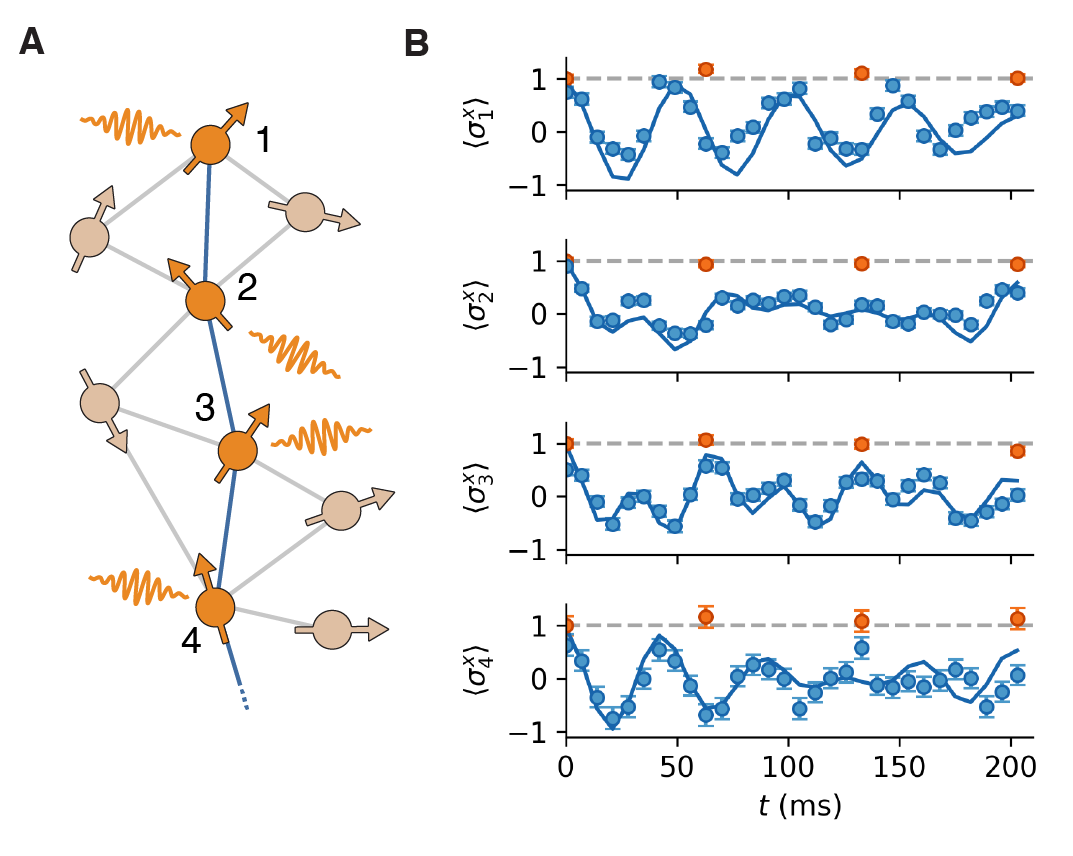}
    \caption{{\bf Isolating spin chains.} ({\bf A}) We test the programming of interacting spin chains for the first 4 spins of the 9-spin chain (Figs. \ref{fig:1Dchain}A,C,D). For $\theta \sim \pi$, the Floquet sequence $U_F$ decouples the spin chain from its environment, but preserves the internal interactions. ({\bf B}) Measured expectation values $\langle \sigma_j^x \rangle$ after initializing the state $\ket{\text{++++}}$ and applying $U_F$ with $\theta = \pi$. Here $t = 2\tau N$ is varied by fixing $\tau = 3.5\,$ms and varying $N$. The blue (orange) points show the evolution with (without) spin-spin interactions \cite{supp}. Blue lines: numerical simulations of only the 4-spin system \cite{supp}. Measurements in this figure and hereafter are corrected for state preparation and measurement errors.}
    \label{fig:xbasis}
\end{figure}

To investigate the DTC phase, we apply a periodic Floquet sequence consisting of free evolution $U_\text{int}(\tau) = \exp(-iH_\text{int}\tau)$, interleaved with global spin rotations $U_x(\theta) = \exp(-i\theta \sum_j^L\sigma_j^x/2)$. To realize the global rotations, we develop multi-frequency rf pulses that simultaneously rotate a chosen subset of spins ($H_\text{rf}$ in Eq. \ref{eq:Hint}) \cite{supp}. We symmetrize the Floquet sequence such that $U_F = [U_\text{int}(\tau)\cdot U_x(\theta) \cdot U_\text{int}(\tau)]^N$, where $N$ is the number of Floquet cycles (Fig. 1B). For $\theta \sim \pi$, this decouples the targeted spins from their environment, while preserving the internal interactions \cite{supp}.

To stabilize MBL under periodic driving, the Hamiltonian must satisfy two requirements \cite{ippoliti2020many, khemani2019brief, yao2017discrete}. First, the spin-spin interactions $J_{jk}$ must be sufficiently short-ranged. For power-law interactions that fall off as $1/r^\alpha$, it is believed that MBL requires $\alpha > 2d$, where $d$ is the dimension of the system \cite{avalanche, yao2014many,burin2015many,ippoliti2020many}. For dipole-dipole interactions, $\alpha = 3$. Because the nuclear spins are randomly positioned in $d=3$ dimensions, the short-ranged requirement is not naturally met. To resolve this, we program an effective 1D spin chain using a subset of 9 spins  (Figs. \ref{fig:1Dchain}A,C \cite{supp}). As a function of site distance across the chain, a fit to the averaged couplings falls of as $1/|j-k|^{2.5(1)}$ (Fig. \ref{fig:1Dchain}D), confirming that the finite-sized chain maps onto an approximately 1D system whose interactions fall off sufficiently fast to be compatible with MBL. Second, since the periodic rotations in $U_F$ approximately cancel the on-site disorder terms $h_j$, the system must exhibit Ising-even disorder to stabilize MBL in the Floquet setting \cite{khemani2019brief,ippoliti2020many,yao2017discrete}. This corresponds to disorder in the couplings $J_{jk}$, which is naturally satisfied here (Fig. \ref{fig:1Dchain}C).

\begin{figure*}
    \centering
    \includegraphics[width=1.0\textwidth]{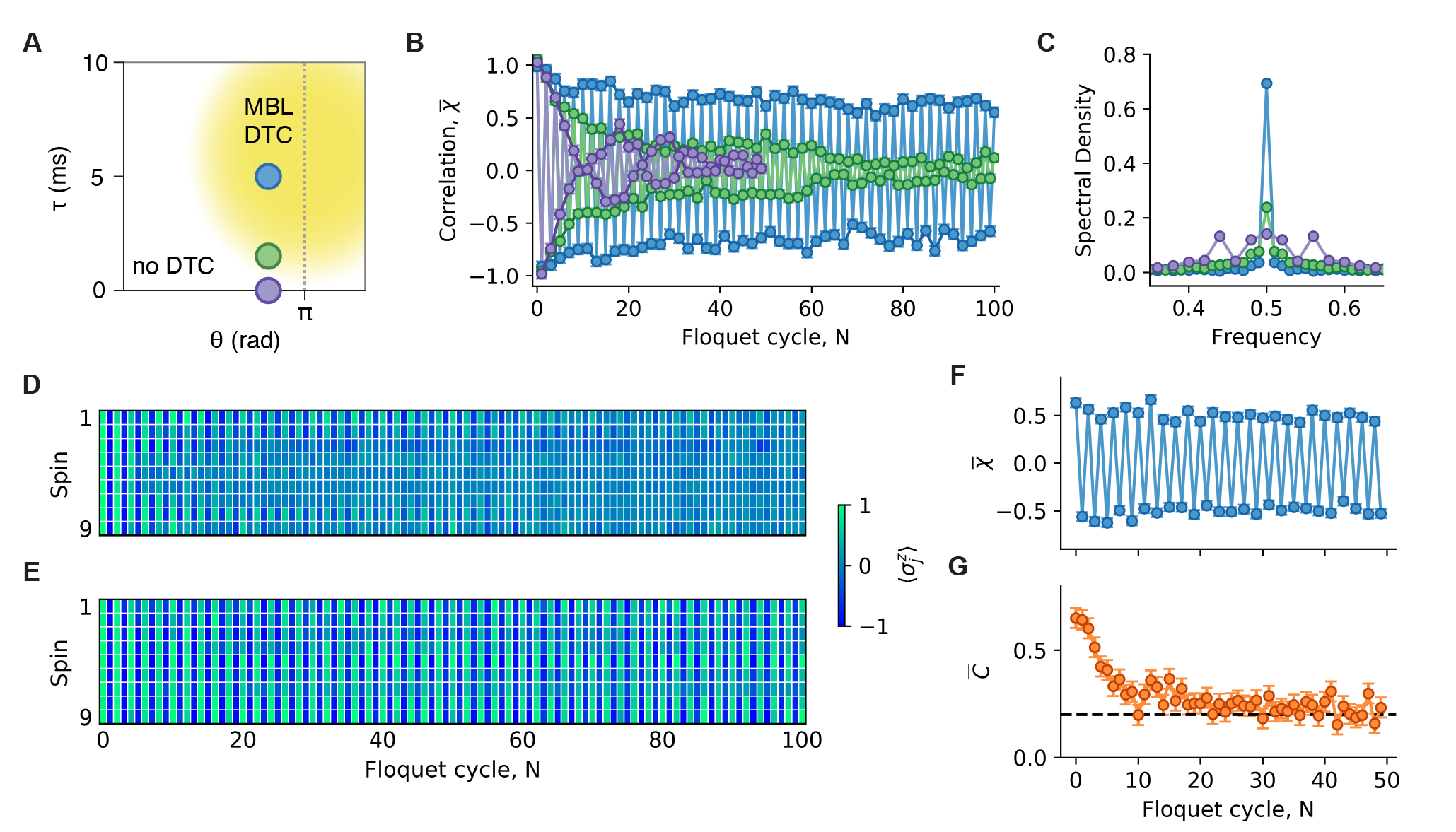}
      \caption{{\bf Discrete time crystal in the 9-spin chain.} ({\bf A}) Sketch of the phase diagram as a function of $\tau$ and $\theta$ when applying $U_F$ 
      (Fig. 1B) \cite{khemani2019brief}. The yellow region indicates the many-body-localized DTC phase. The colored points mark three combinations of \{$\theta$,$\tau$\} that illustrate the DTC phase transition. Additional data for other values are given in the supplementary materials \cite{supp}. ({\bf B}) Averaged two-point correlation $\overline{\chi}$ as a function of the number of Floquet cycles $N$, for $\theta = 0.95\pi$ and initial state $\ket{\u\u\u\u\u\u\u\u\u}$. Without interactions (purple \cite{supp}), $\overline{\chi}$ decays quickly. With small interactions ($\tau = 1.55\,$ms, green), the system is on the edge of the transition to the DTC phase. With strong interactions ($\tau = 5\,$ms, blue), the subharmonic response is stable and persists over all 100 Floquet cycles. ({\bf C}) The corresponding Fourier transforms show a sharp peak at $f = 0.5$ emerging as the system enters the DTC phase. ({\bf D} and {\bf E}) Individual spin expectation values $\langle \sigma_j^z \rangle$ for interaction times $\tau = 1.55\,$ms (D) and $\tau = 5\,$ms (E). ({\bf F} and {\bf G}) Averaged two-point correlation $\overline{\chi}$ (F) and coherence $\overline{C}$ (G) after preparing the superposition state $[\cos(\pi/8)\ket{\u} + \sin(\pi/8)\ket{\d}]^{\otimes 9}$ and applying $U_F$ with $\tau = 5\,$ms. The subharmonic response in $\overline{\chi}$ is preserved, while $\overline{C}$ quickly decays due to interaction-induced local dephasing. The dashed line in (G) indicates a reference value for $\overline{C}$ measured after preparing the state $\ket{\u}^{\otimes 9}$ \cite{supp}.}
    \label{fig:DTC}
\end{figure*}

To reveal the signature spatiotemporal order of the DTC phase, one must prepare a variety of initial states and perform site-resolved measurements \cite{ippoliti2020many}. We use a combination of new and existing methods to realize the required initialization, single-spin control, and individual single-shot measurement for all spins in the chain (Fig. \ref{fig:1Dchain}B).

First, we initialize the spins through a recently introduced dynamical-nuclear-polarization sequence called PulsePol \cite{schwartz2018robust}. This sequence polarizes nuclear spins in the vicinity of the NV center and prepares the 1D chain in the state $\ket{\u\u\u\u\u\u\u\u\u}$. We analyze and optimize the polarization transfer in the supplementary materials \cite{supp}. Subsequently, each spin can be independently rotated to an arbitrary state by selective rf pulses \cite{supp}.

Second, after Floquet evolution, we read out the spins by sequentially mapping their $\langle \sigma_j^z \rangle$ expectation values to the NV electronic spin \cite{supp}, and measuring the electronic-spin state via resonant optical excitation \cite{bradley2019ten}. Spins $j$=2,5,6,8 can be directly accessed using previously developed electron-nuclear two-qubit gates \cite{bradley2019ten}. To access the other spins ($j$=1,3,4,7,9), which couple weakly to the NV, we develop a protocol based on nuclear-nuclear two-qubit gates through spin-echo double resonance \cite{supp}. We use these gates to map the spin states to other, directly accessible, spins in the chain. Fig. \ref{fig:1Dchain}E shows the measured $\langle \sigma_j^z \rangle$ expectation values after preparing the state $\ket{\u\u\u\u\u\u\u\u\u}$.  

We verify that we can isolate the dynamics of a subset of spins by studying the first 4 spins of the 9-spin chain (Fig. 2A). We prepare the superposition state $\ket{\text{++++}}$, where $\ket{+} = (\ket{\u} + \ket{\d})/\sqrt{2}$, and apply $U_F$ with $\theta = \pi$. We first verify that the state is preserved when each spin is individually decoupled to remove interactions (Fig. 2B) \cite{supp}. In contrast, with internal interactions, the four spins entangle and undergo complex dynamics. The measured evolution matches a numerical simulation containing only the 4 spins, indicating that the system is strongly interacting and protected from external decoherence.

With this capability confirmed, we turn to the 9-spin chain and the DTC phase. The expectation for the DTC phase is a long-lived period-doubled response that is stabilized against perturbations of $U_F$ through many-body interactions. To illustrate this, we set $\theta = 0.95 \pi$, a perturbation from the ideal value of $\pi$, and tune the system through the DTC phase transition by changing $\tau$, which effectively sets the interaction strength (see Figs. \ref{fig:DTC}A-C). 

We first investigate the state $\ket{\u\u\u\u\u\u\u\u\u}$ and consider the averaged two-point correlation function $\overline{\chi} = \frac{1}{L}\sum_{j=1}^L \langle \sigma_j^z (N) \rangle \langle \sigma_j^z (0) \rangle$, where $\langle \sigma_j^z(N) \rangle$ is the expectation value at Floquet cycle $N$ for spin $j$. Without interactions, the deliberate under-rotations ($\theta < \pi$), in combination with naturally present noise in the applied control fields, lead to a rapid decay (Figs. \ref{fig:DTC}B,C). By introducing moderate interactions ($\tau = 1.55$ ms), the system is on the edge of the phase transition, and the interactions begin to stabilize the subharmonic response (Figs. \ref{fig:DTC}B,C,D). Finally, for strong interactions ($\tau = 5$ ms), the subharmonic response is stabilized despite the perturbations of $\theta$ (Figs. \ref{fig:DTC}B,C,E). The individual spin measurements confirm that the spins are synchronized and the signature long-lived spatiotemporal response is observed (Fig. \ref{fig:DTC}E). 

\begin{figure*}
    \centering
    \includegraphics[width=0.75\textwidth]{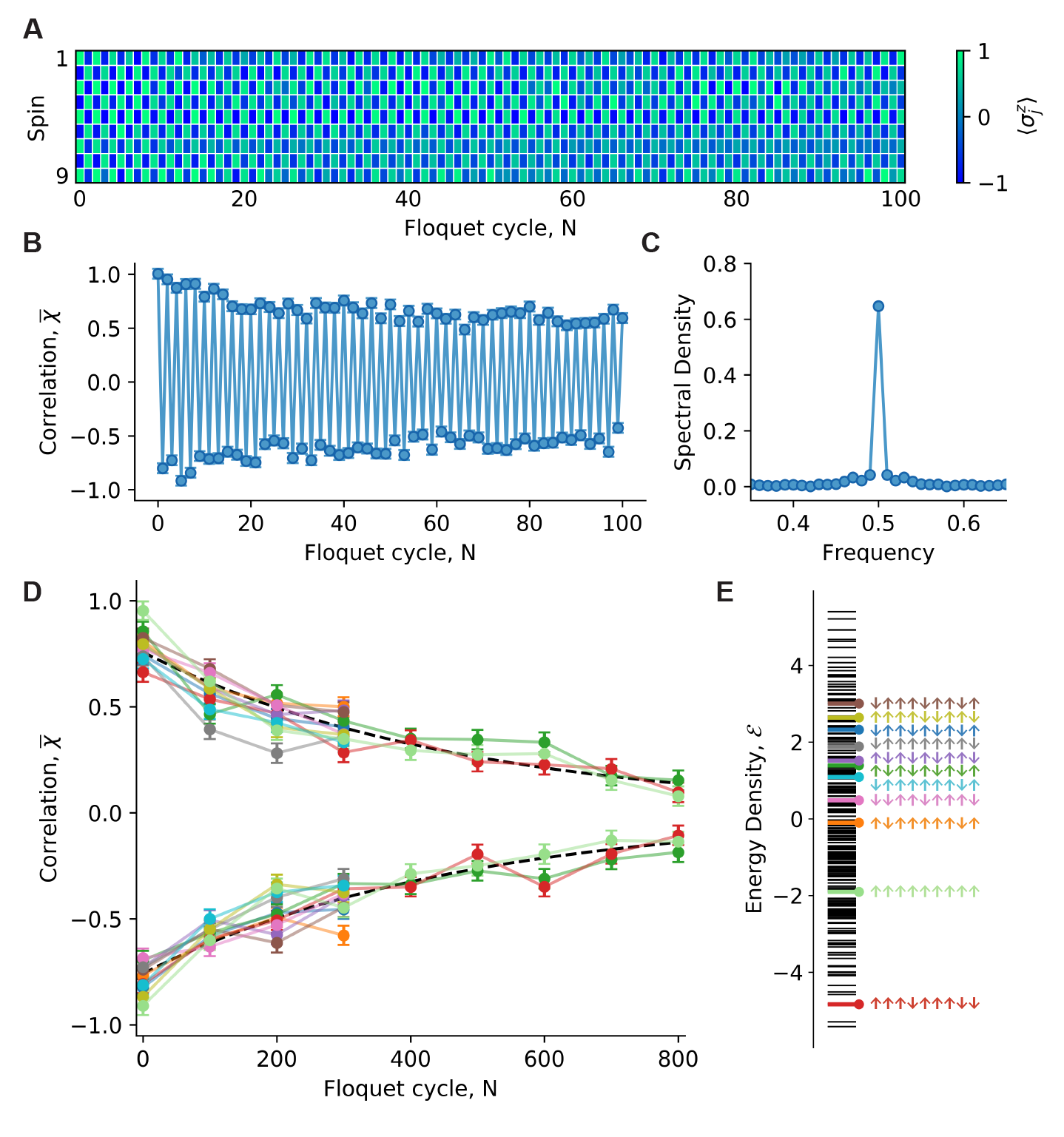}
    \caption{{\bf Observation of the DTC response for generic initial states.} ({\bf A}) Individual spin expectation values $\langle \sigma_j^z \rangle$ as a function of $N$ after initializing the spins in the Néel state $\ket{\u\d\u\d\u\d\u\d\u}$ and applying $U_F$ for $\theta = 0.95\pi$ and $\tau = 5\,$ms. ({\bf B}) Averaged two-point correlation function $\overline{\chi}$, corresponding to the data in (A). The DTC response persists to similar high $N$ as for the polarized state (Fig. \ref{fig:DTC}B). ({\bf C}) Fourier transform of the data in (B), showing the period-doubled response. ({\bf D}) Average correlation for even (upper curve) and odd (lower curve) $N$ for 9 randomly chosen initial states, plus the polarized state and the Néel state (indicated in (E)) with $\theta = 0.95\pi$ and $\tau = 5\,$ms. Each data point is the average over even/odd integers in the range $N$ to $N+10$. Three of the states are measured up to $N = 800$, the others to $N=300$. The dashed black line is a fit of $|\overline{\chi}|$, averaged over the three states measured to $N = 800$, using a phenomenological function $f(N) = Ae^{-N/N_{1/e}}$, giving $A = 0.76(1)$ and $N_{1/e} = 472(17)$. ({\bf E}) Calculated energy density $\mathcal{E}$ for all possible states of the form $\bigotimes_j^L \ket{m_j}$, $m_j \in \{\u,\d\}$ (black lines). The initial states measured in (D) are indicated by the corresponding colors.}
    \label{fig:initstates}
\end{figure*}

To rule out trivial non-interacting explanations, we prepare the superposition state $[\cos(\pi/8)\ket{\u} + \sin(\pi/8)\ket{\d}]^{\otimes 9}$ and perform full single-qubit tomography for each spin for different values of $N$ \cite{else2016floquet}. The two-point correlation $\overline{\chi}$, shows a persistent subharmonic response similar to the initial state $\ket{\u\u\u\u\u\u\u\u\u}$ (Fig. \ref{fig:DTC}F). In contrast, the coherence $\overline{C} = \frac{1}{L}\sum_{j=1}^L \sqrt{\langle \sigma_j^x \rangle^2 + \langle \sigma_j^y \rangle^2}$ shows a quick decay on a timescale of approximately $10$ Floquet cycles, indicating rapid local dephasing due to internal many-body interactions that generate entanglement across the system (Fig. \ref{fig:DTC}G).

While the results shown in Fig. \ref{fig:DTC} are consistent with a DTC, these measurements alone do not distinguish the many-body-localized DTC phase from prethermal responses \cite{khemani2019brief, ippoliti2020many, kyprianidis2021observation}. In particular, the hallmark of the MBL DTC phase is robust spatiotemporal order for generic initial states. Conversely, prethermal responses only exhibit long-lived oscillations for a particular range of initial states \cite{khemani2019brief, ippoliti2020many}.

We study a range of generic initial states of the form $\bigotimes_j^L \ket{m_j}$, $m_j \in \{\u,\d\}$, starting with  the Néel state $\ket{\u\d\u\d\u\d\u\d\u}$ (Fig. \ref{fig:initstates}A). Like the polarized state, the Néel state shows a stable, period-doubled response (Figs. \ref{fig:initstates}B,C). Fig. \ref{fig:initstates}D shows the decay of the DTC response for the Néel state, the polarized state, and a further 9 randomly chosen initial states. To illustrate that a variety of states with different properties are considered, we evaluate their energy density $\mathcal{E} = \langle H_\text{eff} \rangle / J_0 L$ (Fig. 4E), where $J_0$ is the average nearest-neighbor coupling strength (Fig. 1D) and $H_\text{eff}$ is the leading order term in the Floquet-Magnus expansion of $U_F$ \cite{supp}. The selected initial states cover a range of energy densities. The response shows no significant dependence on the initial state up to $N=800$, consistent with a DTC stabilized by MBL. 

To highlight the importance of disorder, we perform a numerical investigation with and without disorder \cite{supp}. For the parameters of the disordered 9-spin chain without decoherence, we find a stable period-doubled response up to $N \sim 10^{6}$ for all initial states. For a hypothetical 9-spin chain without disorder, but with the same average couplings, there is no MBL and the time-crystalline response is state-dependent and fully decays within 300 Floquet cycles for some states. These simulations show that robust spatiotemporal order associated with the many-body-localized DTC phase can be distinguished from disorder-free prethermal responses within the experimentally accessible timescales. 

While the DTC phase in an ideal isolated system is predicted to persist to arbitrary times, environmental decoherence inevitably causes decay in any experimental implementation. We observe a 1/e decay value of $N_{1/e} = 472(17)$ — corresponding to a time $\sim 4.7\,$s — highlighting that our platform is highly isolated. However, understanding how the DTC response is affected by different decoherence mechanisms is an outstanding challenge. While the dominant decoherence mechanism for the spins is dephasing with a timescale on the order of seconds \cite{bradley2019ten}, the DTC phase is expected to be particularly robust to such effects \cite{ippoliti2020many}. This suggests that the observed decay arises from a more subtle interplay between the Floquet sequence and the environment, which might be suppressed using future optimized decoupling sequences. Crucially, the numerical calculations without decoherence show that the finite size of the spin chain does not limit the observed DTC response.


In conclusion, we present an observation of the hallmark signatures of the many-body-localized DTC phase. Unlike previous experiments, our quantum simulator operates in a regime consistent with MBL and the DTC response is observed to be stable for generic initial states. This result highlights the importance of both many-body interactions and disorder for stabilizing time-crystalline order. The developed methods provide new opportunities to investigate Floquet phases of matter, including topologically protected phases \cite{khemani2019brief}, and time-crystalline order in a variety of settings complementary to MBL, such as open systems where the interplay between dissipation and interactions leads to distinct DTC phenomena \cite{gong2018discrete,yao2020classical,lazarides2020time}.

From a broader perspective, this work introduces a programmable quantum simulator based on solid-state spins. By connecting different subsets of spins, larger one-dimensional chains and two- and three-dimensional systems can be realized. The combination of excellent coherence, individual control and site-selective measurement enables the programming of a wide variety of many-body Hamiltonians. Future scalability beyond tens of spins might be achieved by exploiting spins external to the diamond \cite{cai2013large,lovchinsky2017magnetic}, by linking multiple electronic-spin defects through dipolar coupling \cite{dolde2013room}, by photonic remote entanglement \cite{pompili2021realization}, or by combinations of these methods.

\section{Acknowledgments}

We thank W. Hahn and V. V. Dobrovitski for valuable discussions and A. Breitweiser for experimental assistance. {\bf Funding:} This work was supported by
the Netherlands Organisation for Scientific Research
(NWO/OCW) through a Vidi grant and as part of the
Frontiers of Nanoscience (NanoFront) programme and the Quantum Software Consortium programme (Project No. 024.003.037/3368). This
project has received funding from the European Research
Council (ERC) under the European Union’s Horizon
2020 research and innovation programme (grant agreement No. 852410). This project (QIA) has received funding from the European Union’s Horizon 2020 research
and innovation programme under grant agreement No
820445. F. M. acknowledges support from the U.S. Department of Energy, Office of Science, Office of Basic Energy Sciences, Materials Sciences and Engineering Division under Contract No. DE-AC02-05-CH11231``High-Coherence Multilayer Superconducting Structures for Large Scale Qubit Integration and Photonic Transduction program (QIS-LBNL)". N. Y. Y. acknowledges support from DARPA DRINQS program (D18AC00033) and from the Army Research Office (grant no. W911NF2110262). {\bf Author contributions:} JR, CEB, FVvdG, AG and THT devised and performed the experiments, developed theoretical calculations and performed numerical simulations. JR, CEB, FVvdG, AG, FM, NYY and THT analyzed the data. JR, CEB, MHA and THT prepared the experimental apparatus. MM and DJT grew the diamond sample. JR, CEB and THT wrote the manuscript with input from all authors. THT supervised the project. {\bf Competing interests:} The authors declare no competing interests. {\bf Data and materials availability:} The data that support the findings of this study are available from the corresponding author upon request.

\bibliography{Qsim_bib}

\begin{thebibliography}{37}%
\makeatletter
\providecommand \@ifxundefined [1]{%
 \@ifx{#1\undefined}
}%
\providecommand \@ifnum [1]{%
 \ifnum #1\expandafter \@firstoftwo
 \else \expandafter \@secondoftwo
 \fi
}%
\providecommand \@ifx [1]{%
 \ifx #1\expandafter \@firstoftwo
 \else \expandafter \@secondoftwo
 \fi
}%
\providecommand \natexlab [1]{#1}%
\providecommand \enquote  [1]{``#1''}%
\providecommand \bibnamefont  [1]{#1}%
\providecommand \bibfnamefont [1]{#1}%
\providecommand \citenamefont [1]{#1}%
\providecommand \href@noop [0]{\@secondoftwo}%
\providecommand \href [0]{\begingroup \@sanitize@url \@href}%
\providecommand \@href[1]{\@@startlink{#1}\@@href}%
\providecommand \@@href[1]{\endgroup#1\@@endlink}%
\providecommand \@sanitize@url [0]{\catcode `\\12\catcode `\$12\catcode
  `\&12\catcode `\#12\catcode `\^12\catcode `\_12\catcode `\%12\relax}%
\providecommand \@@startlink[1]{}%
\providecommand \@@endlink[0]{}%
\providecommand \url  [0]{\begingroup\@sanitize@url \@url }%
\providecommand \@url [1]{\endgroup\@href {#1}{\urlprefix }}%
\providecommand \urlprefix  [0]{URL }%
\providecommand \Eprint [0]{\href }%
\providecommand \doibase [0]{http://dx.doi.org/}%
\providecommand \selectlanguage [0]{\@gobble}%
\providecommand \bibinfo  [0]{\@secondoftwo}%
\providecommand \bibfield  [0]{\@secondoftwo}%
\providecommand \translation [1]{[#1]}%
\providecommand \BibitemOpen [0]{}%
\providecommand \bibitemStop [0]{}%
\providecommand \bibitemNoStop [0]{.\EOS\space}%
\providecommand \EOS [0]{\spacefactor3000\relax}%
\providecommand \BibitemShut  [1]{\csname bibitem#1\endcsname}%
\let\auto@bib@innerbib\@empty
\bibitem [{\citenamefont {Wilczek}(2012)}]{wilczek2012quantum}%
  \BibitemOpen
  \bibfield  {author} {\bibinfo {author} {\bibfnamefont {F.}~\bibnamefont
  {Wilczek}},\ }\href {\doibase 10.1103/PhysRevLett.109.160401} {\bibfield
  {journal} {\bibinfo  {journal} {Phys. Rev. Lett.}\ }\textbf {\bibinfo
  {volume} {109}},\ \bibinfo {pages} {160401} (\bibinfo {year}
  {2012})}\BibitemShut {NoStop}%
\bibitem [{\citenamefont {Watanabe}\ and\ \citenamefont
  {Oshikawa}(2015)}]{watanabe2015absence}%
  \BibitemOpen
  \bibfield  {author} {\bibinfo {author} {\bibfnamefont {H.}~\bibnamefont
  {Watanabe}}\ and\ \bibinfo {author} {\bibfnamefont {M.}~\bibnamefont
  {Oshikawa}},\ }\href {\doibase 10.1103/PhysRevLett.114.251603} {\bibfield
  {journal} {\bibinfo  {journal} {Phys. Rev. Lett.}\ }\textbf {\bibinfo
  {volume} {114}},\ \bibinfo {pages} {251603} (\bibinfo {year}
  {2015})}\BibitemShut {NoStop}%
\bibitem [{\citenamefont {Else}\ \emph {et~al.}(2016)\citenamefont {Else},
  \citenamefont {Bauer},\ and\ \citenamefont {Nayak}}]{else2016floquet}%
  \BibitemOpen
  \bibfield  {author} {\bibinfo {author} {\bibfnamefont {D.~V.}\ \bibnamefont
  {Else}}, \bibinfo {author} {\bibfnamefont {B.}~\bibnamefont {Bauer}}, \ and\
  \bibinfo {author} {\bibfnamefont {C.}~\bibnamefont {Nayak}},\ }\href@noop {}
  {\bibfield  {journal} {\bibinfo  {journal} {Phys. Rev. Lett.}\ }\textbf
  {\bibinfo {volume} {117}},\ \bibinfo {pages} {090402} (\bibinfo {year}
  {2016})}\BibitemShut {NoStop}%
\bibitem [{\citenamefont {Khemani}\ \emph {et~al.}(2016)\citenamefont
  {Khemani}, \citenamefont {Lazarides}, \citenamefont {Moessner},\ and\
  \citenamefont {Sondhi}}]{khemani2016phase}%
  \BibitemOpen
  \bibfield  {author} {\bibinfo {author} {\bibfnamefont {V.}~\bibnamefont
  {Khemani}}, \bibinfo {author} {\bibfnamefont {A.}~\bibnamefont {Lazarides}},
  \bibinfo {author} {\bibfnamefont {R.}~\bibnamefont {Moessner}}, \ and\
  \bibinfo {author} {\bibfnamefont {S.~L.}\ \bibnamefont {Sondhi}},\ }\href
  {\doibase 10.1103/PhysRevLett.116.250401} {\bibfield  {journal} {\bibinfo
  {journal} {Phys. Rev. Lett.}\ }\textbf {\bibinfo {volume} {116}},\ \bibinfo
  {pages} {250401} (\bibinfo {year} {2016})}\BibitemShut {NoStop}%
\bibitem [{\citenamefont {Yao}\ \emph {et~al.}(2017)\citenamefont {Yao},
  \citenamefont {Potter}, \citenamefont {Potirniche},\ and\ \citenamefont
  {Vishwanath}}]{yao2017discrete}%
  \BibitemOpen
  \bibfield  {author} {\bibinfo {author} {\bibfnamefont {N.~Y.}\ \bibnamefont
  {Yao}}, \bibinfo {author} {\bibfnamefont {A.~C.}\ \bibnamefont {Potter}},
  \bibinfo {author} {\bibfnamefont {I.-D.}\ \bibnamefont {Potirniche}}, \ and\
  \bibinfo {author} {\bibfnamefont {A.}~\bibnamefont {Vishwanath}},\ }\href
  {\doibase 10.1103/PhysRevLett.118.030401} {\bibfield  {journal} {\bibinfo
  {journal} {Phys. Rev. Lett.}\ }\textbf {\bibinfo {volume} {118}},\ \bibinfo
  {pages} {030401} (\bibinfo {year} {2017})}\BibitemShut {NoStop}%
\bibitem [{\citenamefont {Khemani}\ \emph {et~al.}(2019)\citenamefont
  {Khemani}, \citenamefont {Moessner},\ and\ \citenamefont
  {Sondhi}}]{khemani2019brief}%
  \BibitemOpen
  \bibfield  {author} {\bibinfo {author} {\bibfnamefont {V.}~\bibnamefont
  {Khemani}}, \bibinfo {author} {\bibfnamefont {R.}~\bibnamefont {Moessner}}, \
  and\ \bibinfo {author} {\bibfnamefont {S.~L.}\ \bibnamefont {Sondhi}},\
  }\href@noop {} {\bibfield  {journal} {\bibinfo  {journal} {arXiv preprint
  arXiv:1910.10745}\ } (\bibinfo {year} {2019})}\BibitemShut {NoStop}%
\bibitem [{\citenamefont {Else}\ \emph {et~al.}(2020)\citenamefont {Else},
  \citenamefont {Monroe}, \citenamefont {Nayak},\ and\ \citenamefont
  {Yao}}]{else2020discrete}%
  \BibitemOpen
  \bibfield  {author} {\bibinfo {author} {\bibfnamefont {D.~V.}\ \bibnamefont
  {Else}}, \bibinfo {author} {\bibfnamefont {C.}~\bibnamefont {Monroe}},
  \bibinfo {author} {\bibfnamefont {C.}~\bibnamefont {Nayak}}, \ and\ \bibinfo
  {author} {\bibfnamefont {N.~Y.}\ \bibnamefont {Yao}},\ }\href@noop {}
  {\bibfield  {journal} {\bibinfo  {journal} {Annu. Rev. Condens. Matter
  Phys.}\ }\textbf {\bibinfo {volume} {11}},\ \bibinfo {pages} {467} (\bibinfo
  {year} {2020})}\BibitemShut {NoStop}%
\bibitem [{\citenamefont {Abanin}\ \emph {et~al.}(2019)\citenamefont {Abanin},
  \citenamefont {Altman}, \citenamefont {Bloch},\ and\ \citenamefont
  {Serbyn}}]{abanin2019colloquium}%
  \BibitemOpen
  \bibfield  {author} {\bibinfo {author} {\bibfnamefont {D.~A.}\ \bibnamefont
  {Abanin}}, \bibinfo {author} {\bibfnamefont {E.}~\bibnamefont {Altman}},
  \bibinfo {author} {\bibfnamefont {I.}~\bibnamefont {Bloch}}, \ and\ \bibinfo
  {author} {\bibfnamefont {M.}~\bibnamefont {Serbyn}},\ }\href@noop {}
  {\bibfield  {journal} {\bibinfo  {journal} {Rev. Mod. Phys.}\ }\textbf
  {\bibinfo {volume} {91}},\ \bibinfo {pages} {021001} (\bibinfo {year}
  {2019})}\BibitemShut {NoStop}%
\bibitem [{\citenamefont {D’Alessio}\ and\ \citenamefont
  {Rigol}(2014)}]{d2014long}%
  \BibitemOpen
  \bibfield  {author} {\bibinfo {author} {\bibfnamefont {L.}~\bibnamefont
  {D’Alessio}}\ and\ \bibinfo {author} {\bibfnamefont {M.}~\bibnamefont
  {Rigol}},\ }\href@noop {} {\bibfield  {journal} {\bibinfo  {journal} {Phys.
  Rev. X}\ }\textbf {\bibinfo {volume} {4}},\ \bibinfo {pages} {041048}
  (\bibinfo {year} {2014})}\BibitemShut {NoStop}%
\bibitem [{\citenamefont {Zhang}\ \emph {et~al.}(2017)\citenamefont {Zhang},
  \citenamefont {Hess}, \citenamefont {Kyprianidis}, \citenamefont {Becker},
  \citenamefont {Lee}, \citenamefont {Smith}, \citenamefont {Pagano},
  \citenamefont {Potirniche}, \citenamefont {Potter}, \citenamefont
  {Vishwanath} \emph {et~al.}}]{zhang2017observation}%
  \BibitemOpen
  \bibfield  {author} {\bibinfo {author} {\bibfnamefont {J.}~\bibnamefont
  {Zhang}}, \bibinfo {author} {\bibfnamefont {P.}~\bibnamefont {Hess}},
  \bibinfo {author} {\bibfnamefont {A.}~\bibnamefont {Kyprianidis}}, \bibinfo
  {author} {\bibfnamefont {P.}~\bibnamefont {Becker}}, \bibinfo {author}
  {\bibfnamefont {A.}~\bibnamefont {Lee}}, \bibinfo {author} {\bibfnamefont
  {J.}~\bibnamefont {Smith}}, \bibinfo {author} {\bibfnamefont
  {G.}~\bibnamefont {Pagano}}, \bibinfo {author} {\bibfnamefont {I.-D.}\
  \bibnamefont {Potirniche}}, \bibinfo {author} {\bibfnamefont {A.~C.}\
  \bibnamefont {Potter}}, \bibinfo {author} {\bibfnamefont {A.}~\bibnamefont
  {Vishwanath}},  \emph {et~al.},\ }\href@noop {} {\bibfield  {journal}
  {\bibinfo  {journal} {Nature}\ }\textbf {\bibinfo {volume} {543}},\ \bibinfo
  {pages} {217} (\bibinfo {year} {2017})}\BibitemShut {NoStop}%
\bibitem [{\citenamefont {Kyprianidis}\ \emph {et~al.}(2021)\citenamefont
  {Kyprianidis}, \citenamefont {Machado}, \citenamefont {Morong}, \citenamefont
  {Becker}, \citenamefont {Collins}, \citenamefont {Else}, \citenamefont
  {Feng}, \citenamefont {Hess}, \citenamefont {Nayak}, \citenamefont {Pagano}
  \emph {et~al.}}]{kyprianidis2021observation}%
  \BibitemOpen
  \bibfield  {author} {\bibinfo {author} {\bibfnamefont {A.}~\bibnamefont
  {Kyprianidis}}, \bibinfo {author} {\bibfnamefont {F.}~\bibnamefont
  {Machado}}, \bibinfo {author} {\bibfnamefont {W.}~\bibnamefont {Morong}},
  \bibinfo {author} {\bibfnamefont {P.}~\bibnamefont {Becker}}, \bibinfo
  {author} {\bibfnamefont {K.~S.}\ \bibnamefont {Collins}}, \bibinfo {author}
  {\bibfnamefont {D.~V.}\ \bibnamefont {Else}}, \bibinfo {author}
  {\bibfnamefont {L.}~\bibnamefont {Feng}}, \bibinfo {author} {\bibfnamefont
  {P.~W.}\ \bibnamefont {Hess}}, \bibinfo {author} {\bibfnamefont
  {C.}~\bibnamefont {Nayak}}, \bibinfo {author} {\bibfnamefont
  {G.}~\bibnamefont {Pagano}},  \emph {et~al.},\ }\href@noop {} {\bibfield
  {journal} {\bibinfo  {journal} {Science}\ }\textbf {\bibinfo {volume}
  {372}},\ \bibinfo {pages} {1192} (\bibinfo {year} {2021})}\BibitemShut
  {NoStop}%
\bibitem [{\citenamefont {Choi}\ \emph {et~al.}(2017)\citenamefont {Choi},
  \citenamefont {Choi}, \citenamefont {Landig}, \citenamefont {Kucsko},
  \citenamefont {Zhou}, \citenamefont {Isoya}, \citenamefont {Jelezko},
  \citenamefont {Onoda}, \citenamefont {Sumiya}, \citenamefont {Khemani} \emph
  {et~al.}}]{choi2017observation}%
  \BibitemOpen
  \bibfield  {author} {\bibinfo {author} {\bibfnamefont {S.}~\bibnamefont
  {Choi}}, \bibinfo {author} {\bibfnamefont {J.}~\bibnamefont {Choi}}, \bibinfo
  {author} {\bibfnamefont {R.}~\bibnamefont {Landig}}, \bibinfo {author}
  {\bibfnamefont {G.}~\bibnamefont {Kucsko}}, \bibinfo {author} {\bibfnamefont
  {H.}~\bibnamefont {Zhou}}, \bibinfo {author} {\bibfnamefont {J.}~\bibnamefont
  {Isoya}}, \bibinfo {author} {\bibfnamefont {F.}~\bibnamefont {Jelezko}},
  \bibinfo {author} {\bibfnamefont {S.}~\bibnamefont {Onoda}}, \bibinfo
  {author} {\bibfnamefont {H.}~\bibnamefont {Sumiya}}, \bibinfo {author}
  {\bibfnamefont {V.}~\bibnamefont {Khemani}},  \emph {et~al.},\ }\href@noop {}
  {\bibfield  {journal} {\bibinfo  {journal} {Nature}\ }\textbf {\bibinfo
  {volume} {543}},\ \bibinfo {pages} {221} (\bibinfo {year}
  {2017})}\BibitemShut {NoStop}%
\bibitem [{\citenamefont {O'Sullivan}\ \emph {et~al.}(2020)\citenamefont
  {O'Sullivan}, \citenamefont {Lunt}, \citenamefont {Zollitsch}, \citenamefont
  {Thewalt}, \citenamefont {Morton},\ and\ \citenamefont
  {Pal}}]{osullivan2020signatures}%
  \BibitemOpen
  \bibfield  {author} {\bibinfo {author} {\bibfnamefont {J.}~\bibnamefont
  {O'Sullivan}}, \bibinfo {author} {\bibfnamefont {O.}~\bibnamefont {Lunt}},
  \bibinfo {author} {\bibfnamefont {C.~W.}\ \bibnamefont {Zollitsch}}, \bibinfo
  {author} {\bibfnamefont {M.~L.~W.}\ \bibnamefont {Thewalt}}, \bibinfo
  {author} {\bibfnamefont {J.~J.~L.}\ \bibnamefont {Morton}}, \ and\ \bibinfo
  {author} {\bibfnamefont {A.}~\bibnamefont {Pal}},\ }\href {\doibase
  10.1088/1367-2630/ab9fbe} {\bibfield  {journal} {\bibinfo  {journal} {New J.
  Phys.}\ }\textbf {\bibinfo {volume} {22}},\ \bibinfo {pages} {085001}
  (\bibinfo {year} {2020})}\BibitemShut {NoStop}%
\bibitem [{\citenamefont {Rovny}\ \emph {et~al.}(2018)\citenamefont {Rovny},
  \citenamefont {Blum},\ and\ \citenamefont {Barrett}}]{rovny2018observation}%
  \BibitemOpen
  \bibfield  {author} {\bibinfo {author} {\bibfnamefont {J.}~\bibnamefont
  {Rovny}}, \bibinfo {author} {\bibfnamefont {R.~L.}\ \bibnamefont {Blum}}, \
  and\ \bibinfo {author} {\bibfnamefont {S.~E.}\ \bibnamefont {Barrett}},\
  }\href@noop {} {\bibfield  {journal} {\bibinfo  {journal} {Phys. Rev. Lett.}\
  }\textbf {\bibinfo {volume} {120}},\ \bibinfo {pages} {180603} (\bibinfo
  {year} {2018})}\BibitemShut {NoStop}%
\bibitem [{\citenamefont {Pal}\ \emph {et~al.}(2018)\citenamefont {Pal},
  \citenamefont {Nishad}, \citenamefont {Mahesh},\ and\ \citenamefont
  {Sreejith}}]{pal2018temporal}%
  \BibitemOpen
  \bibfield  {author} {\bibinfo {author} {\bibfnamefont {S.}~\bibnamefont
  {Pal}}, \bibinfo {author} {\bibfnamefont {N.}~\bibnamefont {Nishad}},
  \bibinfo {author} {\bibfnamefont {T.~S.}\ \bibnamefont {Mahesh}}, \ and\
  \bibinfo {author} {\bibfnamefont {G.~J.}\ \bibnamefont {Sreejith}},\
  }\href@noop {} {\bibfield  {journal} {\bibinfo  {journal} {Phys. Rev. Lett.}\
  }\textbf {\bibinfo {volume} {120}},\ \bibinfo {pages} {180602} (\bibinfo
  {year} {2018})}\BibitemShut {NoStop}%
\bibitem [{\citenamefont {Smits}\ \emph {et~al.}(2018)\citenamefont {Smits},
  \citenamefont {Liao}, \citenamefont {Stoof},\ and\ \citenamefont {van~der
  Straten}}]{smits2018observation}%
  \BibitemOpen
  \bibfield  {author} {\bibinfo {author} {\bibfnamefont {J.}~\bibnamefont
  {Smits}}, \bibinfo {author} {\bibfnamefont {L.}~\bibnamefont {Liao}},
  \bibinfo {author} {\bibfnamefont {H.~T.~C.}\ \bibnamefont {Stoof}}, \ and\
  \bibinfo {author} {\bibfnamefont {P.}~\bibnamefont {van~der Straten}},\
  }\href {\doibase 10.1103/PhysRevLett.121.185301} {\bibfield  {journal}
  {\bibinfo  {journal} {Phys. Rev. Lett.}\ }\textbf {\bibinfo {volume} {121}},\
  \bibinfo {pages} {185301} (\bibinfo {year} {2018})}\BibitemShut {NoStop}%
\bibitem [{\citenamefont {Autti}\ \emph {et~al.}(2018)\citenamefont {Autti},
  \citenamefont {Eltsov},\ and\ \citenamefont
  {Volovik}}]{autti2018observation}%
  \BibitemOpen
  \bibfield  {author} {\bibinfo {author} {\bibfnamefont {S.}~\bibnamefont
  {Autti}}, \bibinfo {author} {\bibfnamefont {V.~B.}\ \bibnamefont {Eltsov}}, \
  and\ \bibinfo {author} {\bibfnamefont {G.~E.}\ \bibnamefont {Volovik}},\
  }\href {\doibase 10.1103/PhysRevLett.120.215301} {\bibfield  {journal}
  {\bibinfo  {journal} {Phys. Rev. Lett.}\ }\textbf {\bibinfo {volume} {120}},\
  \bibinfo {pages} {215301} (\bibinfo {year} {2018})}\BibitemShut {NoStop}%
\bibitem [{\citenamefont {Frey}\ and\ \citenamefont
  {Rachel}(2021)}]{frey2021simulating}%
  \BibitemOpen
  \bibfield  {author} {\bibinfo {author} {\bibfnamefont {P.}~\bibnamefont
  {Frey}}\ and\ \bibinfo {author} {\bibfnamefont {S.}~\bibnamefont {Rachel}},\
  }\href@noop {} {\bibfield  {journal} {\bibinfo  {journal} {arXiv preprint
  arXiv:2105.06632}\ } (\bibinfo {year} {2021})}\BibitemShut {NoStop}%
\bibitem [{\citenamefont {Ippoliti}\ \emph {et~al.}(2020)\citenamefont
  {Ippoliti}, \citenamefont {Kechedzhi}, \citenamefont {Moessner},
  \citenamefont {Sondhi},\ and\ \citenamefont {Khemani}}]{ippoliti2020many}%
  \BibitemOpen
  \bibfield  {author} {\bibinfo {author} {\bibfnamefont {M.}~\bibnamefont
  {Ippoliti}}, \bibinfo {author} {\bibfnamefont {K.}~\bibnamefont {Kechedzhi}},
  \bibinfo {author} {\bibfnamefont {R.}~\bibnamefont {Moessner}}, \bibinfo
  {author} {\bibfnamefont {S.~L.}\ \bibnamefont {Sondhi}}, \ and\ \bibinfo
  {author} {\bibfnamefont {V.}~\bibnamefont {Khemani}},\ }\href@noop {}
  {\bibfield  {journal} {\bibinfo  {journal} {arXiv preprint arXiv:2007.11602}\
  } (\bibinfo {year} {2020})}\BibitemShut {NoStop}%
\bibitem [{\citenamefont {Else}\ \emph {et~al.}(2017)\citenamefont {Else},
  \citenamefont {Bauer},\ and\ \citenamefont {Nayak}}]{else2017prethermal}%
  \BibitemOpen
  \bibfield  {author} {\bibinfo {author} {\bibfnamefont {D.~V.}\ \bibnamefont
  {Else}}, \bibinfo {author} {\bibfnamefont {B.}~\bibnamefont {Bauer}}, \ and\
  \bibinfo {author} {\bibfnamefont {C.}~\bibnamefont {Nayak}},\ }\href@noop {}
  {\bibfield  {journal} {\bibinfo  {journal} {Phys. Rev. X}\ }\textbf {\bibinfo
  {volume} {7}},\ \bibinfo {pages} {011026} (\bibinfo {year}
  {2017})}\BibitemShut {NoStop}%
\bibitem [{\citenamefont {Machado}\ \emph {et~al.}(2020)\citenamefont
  {Machado}, \citenamefont {Else}, \citenamefont {Kahanamoku-Meyer},
  \citenamefont {Nayak},\ and\ \citenamefont {Yao}}]{machado2020long}%
  \BibitemOpen
  \bibfield  {author} {\bibinfo {author} {\bibfnamefont {F.}~\bibnamefont
  {Machado}}, \bibinfo {author} {\bibfnamefont {D.~V.}\ \bibnamefont {Else}},
  \bibinfo {author} {\bibfnamefont {G.~D.}\ \bibnamefont {Kahanamoku-Meyer}},
  \bibinfo {author} {\bibfnamefont {C.}~\bibnamefont {Nayak}}, \ and\ \bibinfo
  {author} {\bibfnamefont {N.~Y.}\ \bibnamefont {Yao}},\ }\href@noop {}
  {\bibfield  {journal} {\bibinfo  {journal} {Phys. Rev. X}\ }\textbf {\bibinfo
  {volume} {10}},\ \bibinfo {pages} {011043} (\bibinfo {year}
  {2020})}\BibitemShut {NoStop}%
\bibitem [{\citenamefont {Peng}\ \emph {et~al.}(2021)\citenamefont {Peng},
  \citenamefont {Yin}, \citenamefont {Huang}, \citenamefont {Ramanathan},\ and\
  \citenamefont {Cappellaro}}]{peng2021floquet}%
  \BibitemOpen
  \bibfield  {author} {\bibinfo {author} {\bibfnamefont {P.}~\bibnamefont
  {Peng}}, \bibinfo {author} {\bibfnamefont {C.}~\bibnamefont {Yin}}, \bibinfo
  {author} {\bibfnamefont {X.}~\bibnamefont {Huang}}, \bibinfo {author}
  {\bibfnamefont {C.}~\bibnamefont {Ramanathan}}, \ and\ \bibinfo {author}
  {\bibfnamefont {P.}~\bibnamefont {Cappellaro}},\ }\href@noop {} {\bibfield
  {journal} {\bibinfo  {journal} {Nat. Phys.}\ }\textbf {\bibinfo {volume}
  {17}},\ \bibinfo {pages} {444} (\bibinfo {year} {2021})}\BibitemShut
  {NoStop}%
\bibitem [{\citenamefont {Abobeih}\ \emph {et~al.}(2019)\citenamefont
  {Abobeih}, \citenamefont {Randall}, \citenamefont {Bradley}, \citenamefont
  {Bartling}, \citenamefont {Bakker}, \citenamefont {Degen}, \citenamefont
  {Markham}, \citenamefont {Twitchen},\ and\ \citenamefont
  {Taminiau}}]{abobeih2019atomic}%
  \BibitemOpen
  \bibfield  {author} {\bibinfo {author} {\bibfnamefont {M.~H.}\ \bibnamefont
  {Abobeih}}, \bibinfo {author} {\bibfnamefont {J.}~\bibnamefont {Randall}},
  \bibinfo {author} {\bibfnamefont {C.~E.}\ \bibnamefont {Bradley}}, \bibinfo
  {author} {\bibfnamefont {H.~P.}\ \bibnamefont {Bartling}}, \bibinfo {author}
  {\bibfnamefont {M.~A.}\ \bibnamefont {Bakker}}, \bibinfo {author}
  {\bibfnamefont {M.~J.}\ \bibnamefont {Degen}}, \bibinfo {author}
  {\bibfnamefont {M.}~\bibnamefont {Markham}}, \bibinfo {author} {\bibfnamefont
  {D.~J.}\ \bibnamefont {Twitchen}}, \ and\ \bibinfo {author} {\bibfnamefont
  {T.~H.}\ \bibnamefont {Taminiau}},\ }\href@noop {} {\bibfield  {journal}
  {\bibinfo  {journal} {Nature}\ }\textbf {\bibinfo {volume} {576}},\ \bibinfo
  {pages} {411} (\bibinfo {year} {2019})}\BibitemShut {NoStop}%
\bibitem [{\citenamefont {Schwartz}\ \emph {et~al.}(2018)\citenamefont
  {Schwartz}, \citenamefont {Scheuer}, \citenamefont {Tratzmiller},
  \citenamefont {M{\"u}ller}, \citenamefont {Chen}, \citenamefont {Dhand},
  \citenamefont {Wang}, \citenamefont {M{\"u}ller}, \citenamefont {Naydenov},
  \citenamefont {Jelezko} \emph {et~al.}}]{schwartz2018robust}%
  \BibitemOpen
  \bibfield  {author} {\bibinfo {author} {\bibfnamefont {I.}~\bibnamefont
  {Schwartz}}, \bibinfo {author} {\bibfnamefont {J.}~\bibnamefont {Scheuer}},
  \bibinfo {author} {\bibfnamefont {B.}~\bibnamefont {Tratzmiller}}, \bibinfo
  {author} {\bibfnamefont {S.}~\bibnamefont {M{\"u}ller}}, \bibinfo {author}
  {\bibfnamefont {Q.}~\bibnamefont {Chen}}, \bibinfo {author} {\bibfnamefont
  {I.}~\bibnamefont {Dhand}}, \bibinfo {author} {\bibfnamefont {Z.-Y.}\
  \bibnamefont {Wang}}, \bibinfo {author} {\bibfnamefont {C.}~\bibnamefont
  {M{\"u}ller}}, \bibinfo {author} {\bibfnamefont {B.}~\bibnamefont
  {Naydenov}}, \bibinfo {author} {\bibfnamefont {F.}~\bibnamefont {Jelezko}},
  \emph {et~al.},\ }\href@noop {} {\bibfield  {journal} {\bibinfo  {journal}
  {Sci. Adv.}\ }\textbf {\bibinfo {volume} {4}},\ \bibinfo {pages} {eaat8978}
  (\bibinfo {year} {2018})}\BibitemShut {NoStop}%
\bibitem [{sup()}]{supp}%
  \BibitemOpen
  \href@noop {} {\bibinfo  {journal} {See supplementary materials}\
  }\BibitemShut {NoStop}%
\bibitem [{\citenamefont {Bradley}\ \emph {et~al.}(2019)\citenamefont
  {Bradley}, \citenamefont {Randall}, \citenamefont {Abobeih}, \citenamefont
  {Berrevoets}, \citenamefont {Degen}, \citenamefont {Bakker}, \citenamefont
  {Markham}, \citenamefont {Twitchen},\ and\ \citenamefont
  {Taminiau}}]{bradley2019ten}%
  \BibitemOpen
\bibfield  {journal} {  }\bibfield  {author} {\bibinfo {author} {\bibfnamefont
  {C.~E.}\ \bibnamefont {Bradley}}, \bibinfo {author} {\bibfnamefont
  {J.}~\bibnamefont {Randall}}, \bibinfo {author} {\bibfnamefont {M.~H.}\
  \bibnamefont {Abobeih}}, \bibinfo {author} {\bibfnamefont {R.~C.}\
  \bibnamefont {Berrevoets}}, \bibinfo {author} {\bibfnamefont {M.~J.}\
  \bibnamefont {Degen}}, \bibinfo {author} {\bibfnamefont {M.~A.}\ \bibnamefont
  {Bakker}}, \bibinfo {author} {\bibfnamefont {M.}~\bibnamefont {Markham}},
  \bibinfo {author} {\bibfnamefont {D.~J.}\ \bibnamefont {Twitchen}}, \ and\
  \bibinfo {author} {\bibfnamefont {T.~H.}\ \bibnamefont {Taminiau}},\
  }\href@noop {} {\bibfield  {journal} {\bibinfo  {journal} {Phys. Rev. X}\
  }\textbf {\bibinfo {volume} {9}},\ \bibinfo {pages} {031045} (\bibinfo {year}
  {2019})}\BibitemShut {NoStop}%
\bibitem [{ava()}]{avalanche}%
  \BibitemOpen
  \href@noop {} {\bibinfo  {journal} {We note that avalanche instabilities can
  also destabilize MBL in power-law interacting systems, and understanding the
  role of such processes is an important outstanding question
  \cite{roeck2017stability}}\ }\BibitemShut {NoStop}%
\bibitem [{\citenamefont {Yao}\ \emph {et~al.}(2014)\citenamefont {Yao},
  \citenamefont {Laumann}, \citenamefont {Gopalakrishnan}, \citenamefont
  {Knap}, \citenamefont {M\"uller}, \citenamefont {Demler},\ and\ \citenamefont
  {Lukin}}]{yao2014many}%
  \BibitemOpen
\bibfield  {journal} {  }\bibfield  {author} {\bibinfo {author} {\bibfnamefont
  {N.~Y.}\ \bibnamefont {Yao}}, \bibinfo {author} {\bibfnamefont {C.~R.}\
  \bibnamefont {Laumann}}, \bibinfo {author} {\bibfnamefont {S.}~\bibnamefont
  {Gopalakrishnan}}, \bibinfo {author} {\bibfnamefont {M.}~\bibnamefont
  {Knap}}, \bibinfo {author} {\bibfnamefont {M.}~\bibnamefont {M\"uller}},
  \bibinfo {author} {\bibfnamefont {E.~A.}\ \bibnamefont {Demler}}, \ and\
  \bibinfo {author} {\bibfnamefont {M.~D.}\ \bibnamefont {Lukin}},\ }\href
  {\doibase 10.1103/PhysRevLett.113.243002} {\bibfield  {journal} {\bibinfo
  {journal} {Phys. Rev. Lett.}\ }\textbf {\bibinfo {volume} {113}},\ \bibinfo
  {pages} {243002} (\bibinfo {year} {2014})}\BibitemShut {NoStop}%
\bibitem [{\citenamefont {Burin}(2015)}]{burin2015many}%
  \BibitemOpen
  \bibfield  {author} {\bibinfo {author} {\bibfnamefont {A.~L.}\ \bibnamefont
  {Burin}},\ }\href {\doibase 10.1103/PhysRevB.91.094202} {\bibfield  {journal}
  {\bibinfo  {journal} {Phys. Rev. B}\ }\textbf {\bibinfo {volume} {91}},\
  \bibinfo {pages} {094202} (\bibinfo {year} {2015})}\BibitemShut {NoStop}%
\bibitem [{\citenamefont {Gong}\ \emph {et~al.}(2018)\citenamefont {Gong},
  \citenamefont {Hamazaki},\ and\ \citenamefont {Ueda}}]{gong2018discrete}%
  \BibitemOpen
  \bibfield  {author} {\bibinfo {author} {\bibfnamefont {Z.}~\bibnamefont
  {Gong}}, \bibinfo {author} {\bibfnamefont {R.}~\bibnamefont {Hamazaki}}, \
  and\ \bibinfo {author} {\bibfnamefont {M.}~\bibnamefont {Ueda}},\ }\href@noop
  {} {\bibfield  {journal} {\bibinfo  {journal} {Phys. Rev. Lett.}\ }\textbf
  {\bibinfo {volume} {120}},\ \bibinfo {pages} {040404} (\bibinfo {year}
  {2018})}\BibitemShut {NoStop}%
\bibitem [{\citenamefont {Yao}\ \emph {et~al.}(2020)\citenamefont {Yao},
  \citenamefont {Nayak}, \citenamefont {Balents},\ and\ \citenamefont
  {Zaletel}}]{yao2020classical}%
  \BibitemOpen
  \bibfield  {author} {\bibinfo {author} {\bibfnamefont {N.~Y.}\ \bibnamefont
  {Yao}}, \bibinfo {author} {\bibfnamefont {C.}~\bibnamefont {Nayak}}, \bibinfo
  {author} {\bibfnamefont {L.}~\bibnamefont {Balents}}, \ and\ \bibinfo
  {author} {\bibfnamefont {M.~P.}\ \bibnamefont {Zaletel}},\ }\href@noop {}
  {\bibfield  {journal} {\bibinfo  {journal} {Nat. Phys.}\ }\textbf {\bibinfo
  {volume} {16}},\ \bibinfo {pages} {438} (\bibinfo {year} {2020})}\BibitemShut
  {NoStop}%
\bibitem [{\citenamefont {Lazarides}\ \emph {et~al.}(2020)\citenamefont
  {Lazarides}, \citenamefont {Roy}, \citenamefont {Piazza},\ and\ \citenamefont
  {Moessner}}]{lazarides2020time}%
  \BibitemOpen
  \bibfield  {author} {\bibinfo {author} {\bibfnamefont {A.}~\bibnamefont
  {Lazarides}}, \bibinfo {author} {\bibfnamefont {S.}~\bibnamefont {Roy}},
  \bibinfo {author} {\bibfnamefont {F.}~\bibnamefont {Piazza}}, \ and\ \bibinfo
  {author} {\bibfnamefont {R.}~\bibnamefont {Moessner}},\ }\href@noop {}
  {\bibfield  {journal} {\bibinfo  {journal} {Phys. Rev. Res.}\ }\textbf
  {\bibinfo {volume} {2}},\ \bibinfo {pages} {022002} (\bibinfo {year}
  {2020})}\BibitemShut {NoStop}%
\bibitem [{\citenamefont {Cai}\ \emph {et~al.}(2013)\citenamefont {Cai},
  \citenamefont {Retzker}, \citenamefont {Jelezko},\ and\ \citenamefont
  {Plenio}}]{cai2013large}%
  \BibitemOpen
  \bibfield  {author} {\bibinfo {author} {\bibfnamefont {J.}~\bibnamefont
  {Cai}}, \bibinfo {author} {\bibfnamefont {A.}~\bibnamefont {Retzker}},
  \bibinfo {author} {\bibfnamefont {F.}~\bibnamefont {Jelezko}}, \ and\
  \bibinfo {author} {\bibfnamefont {M.~B.}\ \bibnamefont {Plenio}},\
  }\href@noop {} {\bibfield  {journal} {\bibinfo  {journal} {Nat. Phys.}\
  }\textbf {\bibinfo {volume} {9}},\ \bibinfo {pages} {168} (\bibinfo {year}
  {2013})}\BibitemShut {NoStop}%
\bibitem [{\citenamefont {Lovchinsky}\ \emph {et~al.}(2017)\citenamefont
  {Lovchinsky}, \citenamefont {Sanchez-Yamagishi}, \citenamefont {Urbach},
  \citenamefont {Choi}, \citenamefont {Fang}, \citenamefont {Andersen},
  \citenamefont {Watanabe}, \citenamefont {Taniguchi}, \citenamefont
  {Bylinskii}, \citenamefont {Kaxiras} \emph
  {et~al.}}]{lovchinsky2017magnetic}%
  \BibitemOpen
  \bibfield  {author} {\bibinfo {author} {\bibfnamefont {I.}~\bibnamefont
  {Lovchinsky}}, \bibinfo {author} {\bibfnamefont {J.~D.}\ \bibnamefont
  {Sanchez-Yamagishi}}, \bibinfo {author} {\bibfnamefont {E.~K.}\ \bibnamefont
  {Urbach}}, \bibinfo {author} {\bibfnamefont {S.}~\bibnamefont {Choi}},
  \bibinfo {author} {\bibfnamefont {S.}~\bibnamefont {Fang}}, \bibinfo {author}
  {\bibfnamefont {T.~I.}\ \bibnamefont {Andersen}}, \bibinfo {author}
  {\bibfnamefont {K.}~\bibnamefont {Watanabe}}, \bibinfo {author}
  {\bibfnamefont {T.}~\bibnamefont {Taniguchi}}, \bibinfo {author}
  {\bibfnamefont {A.}~\bibnamefont {Bylinskii}}, \bibinfo {author}
  {\bibfnamefont {E.}~\bibnamefont {Kaxiras}},  \emph {et~al.},\ }\href@noop {}
  {\bibfield  {journal} {\bibinfo  {journal} {Science}\ }\textbf {\bibinfo
  {volume} {355}},\ \bibinfo {pages} {503} (\bibinfo {year}
  {2017})}\BibitemShut {NoStop}%
\bibitem [{\citenamefont {Dolde}\ \emph {et~al.}(2013)\citenamefont {Dolde},
  \citenamefont {Jakobi}, \citenamefont {Naydenov}, \citenamefont {Zhao},
  \citenamefont {Pezzagna}, \citenamefont {Trautmann}, \citenamefont {Meijer},
  \citenamefont {Neumann}, \citenamefont {Jelezko},\ and\ \citenamefont
  {Wrachtrup}}]{dolde2013room}%
  \BibitemOpen
  \bibfield  {author} {\bibinfo {author} {\bibfnamefont {F.}~\bibnamefont
  {Dolde}}, \bibinfo {author} {\bibfnamefont {I.}~\bibnamefont {Jakobi}},
  \bibinfo {author} {\bibfnamefont {B.}~\bibnamefont {Naydenov}}, \bibinfo
  {author} {\bibfnamefont {N.}~\bibnamefont {Zhao}}, \bibinfo {author}
  {\bibfnamefont {S.}~\bibnamefont {Pezzagna}}, \bibinfo {author}
  {\bibfnamefont {C.}~\bibnamefont {Trautmann}}, \bibinfo {author}
  {\bibfnamefont {J.}~\bibnamefont {Meijer}}, \bibinfo {author} {\bibfnamefont
  {P.}~\bibnamefont {Neumann}}, \bibinfo {author} {\bibfnamefont
  {F.}~\bibnamefont {Jelezko}}, \ and\ \bibinfo {author} {\bibfnamefont
  {J.}~\bibnamefont {Wrachtrup}},\ }\href@noop {} {\bibfield  {journal}
  {\bibinfo  {journal} {Nat. Phys.}\ }\textbf {\bibinfo {volume} {9}},\
  \bibinfo {pages} {139} (\bibinfo {year} {2013})}\BibitemShut {NoStop}%
\bibitem [{\citenamefont {Pompili}\ \emph {et~al.}(2021)\citenamefont
  {Pompili}, \citenamefont {Hermans}, \citenamefont {Baier}, \citenamefont
  {Beukers}, \citenamefont {Humphreys}, \citenamefont {Schouten}, \citenamefont
  {Vermeulen}, \citenamefont {Tiggelman}, \citenamefont {dos Santos~Martins},
  \citenamefont {Dirkse}, \citenamefont {Wehner},\ and\ \citenamefont
  {Hanson}}]{pompili2021realization}%
  \BibitemOpen
  \bibfield  {author} {\bibinfo {author} {\bibfnamefont {M.}~\bibnamefont
  {Pompili}}, \bibinfo {author} {\bibfnamefont {S.~L.~N.}\ \bibnamefont
  {Hermans}}, \bibinfo {author} {\bibfnamefont {S.}~\bibnamefont {Baier}},
  \bibinfo {author} {\bibfnamefont {H.~K.~C.}\ \bibnamefont {Beukers}},
  \bibinfo {author} {\bibfnamefont {P.~C.}\ \bibnamefont {Humphreys}}, \bibinfo
  {author} {\bibfnamefont {R.~N.}\ \bibnamefont {Schouten}}, \bibinfo {author}
  {\bibfnamefont {R.~F.~L.}\ \bibnamefont {Vermeulen}}, \bibinfo {author}
  {\bibfnamefont {M.~J.}\ \bibnamefont {Tiggelman}}, \bibinfo {author}
  {\bibfnamefont {L.}~\bibnamefont {dos Santos~Martins}}, \bibinfo {author}
  {\bibfnamefont {B.}~\bibnamefont {Dirkse}}, \bibinfo {author} {\bibfnamefont
  {S.}~\bibnamefont {Wehner}}, \ and\ \bibinfo {author} {\bibfnamefont
  {R.}~\bibnamefont {Hanson}},\ }\href {\doibase 10.1126/science.abg1919}
  {\bibfield  {journal} {\bibinfo  {journal} {Science}\ }\textbf {\bibinfo
  {volume} {372}},\ \bibinfo {pages} {259} (\bibinfo {year}
  {2021})}\BibitemShut {NoStop}%
\bibitem [{\citenamefont {De~Roeck}\ and\ \citenamefont
  {Huveneers}(2017)}]{roeck2017stability}%
  \BibitemOpen
  \bibfield  {author} {\bibinfo {author} {\bibfnamefont {W.}~\bibnamefont
  {De~Roeck}}\ and\ \bibinfo {author} {\bibfnamefont {F.}~\bibnamefont
  {Huveneers}},\ }\href {\doibase 10.1103/PhysRevB.95.155129} {\bibfield
  {journal} {\bibinfo  {journal} {Phys. Rev. B}\ }\textbf {\bibinfo {volume}
  {95}},\ \bibinfo {pages} {155129} (\bibinfo {year} {2017})}\BibitemShut
  {NoStop}%
\end{thebibliography}%

\end{document}